\documentclass{Interspeech}
\newcommand{\class}[1]{\small \texttt{#1}}
\usepackage{graphicx}
\usepackage{float}

\interspeechcameraready

\title{Challenges in Automated Processing of Speech from Child Wearables: \\ The Case of Voice Type Classifier}

\author[affiliation={1}]{Tarek}{Kunze}
\author[affiliation={1}]{Marianne}{Métais}
\author[affiliation={1}]{Hadrien}{Titeux}
\author[affiliation={1}]{Lucas}{Elbert}
\author[affiliation={1}]{Joseph}{Coffey}
\author[affiliation={1}]{Emmanuel}{Dupoux}
\author[affiliation={1}]{Alejandrina}{Cristia}
\author[affiliation={2}]{Marvin}{Lavechin}

\affiliation{LSCP}{DEC, ENS, EHESS, CNRS, PSL University}{France}
\affiliation{Computational Psycholinguistics Lab.}{Massachusetts Institute of Technology}{United States}
\email{tarek.kunze@ens.psl.eu}

\usepackage{comment}

\begin{document}
\maketitle

\begin{abstract}
    
\noindent Recordings gathered with child-worn devices promised to revolutionize both fundamental and applied speech sciences by allowing the effortless capture of children's naturalistic speech environment and language production. This promise hinges on speech technologies that can transform the sheer mounds of data thus collected into usable information. This paper demonstrates several obstacles blocking progress by summarizing three years' worth of experiments aimed at improving one fundamental task: Voice Type Classification. Our experiments suggest that improvements in representation features, architecture, and parameter search contribute to only marginal gains in performance. More progress is made by focusing on data relevance and quantity, which highlights the importance of collecting data with appropriate permissions to allow sharing. 
\end{abstract}

\section{Introduction}

Child-worn recording devices have transformed both fundamental and applied speech sciences, capturing egocentric audio (or audio-video) that represents speech from the child’s perspective. They enable effortless, long-duration recordings, offering unprecedented insights into children’s language environments \cite{lavechin2022reverse}. Unlike laboratory studies, which may misrepresent everyday speech, or short-form recordings, which are more prone to observer bias, long-form recordings provide more naturalistic data \cite{bergelson2019day}. These recordings are key to reverse-engineering language acquisition \cite{lavechin2022reverse,dupoux2018cognitive} and have also been used in applied research, such as evaluating interventions to enhance parental communication and even providing feedback to parents \cite{weber2017cultural,list2021shifting}.

The value of this technique hinges on the capabilities of advanced speech technologies, which are essential to transform the vast mounds of data collected via child-worn devices into intelligible descriptors usable by researchers and clinicians. Many users of this technology hope for interpretable metrics, such as how much parents talked to and around the child ~\cite{rasanen2021alice}, how advanced the child's vocalizations were~\cite{cychosz2021vocal}, and, in an ideal world, what children and caregivers were talking about ~\cite{lavechin2023brouhaha,sun2024said,long2024babyview}. From the speech technology viewpoint, this results in users' requests for algorithms that accurately identify, categorize, and transcribe verbal interactions assigned to different speakers. Where are we in delivering on this promise?

For the last two decades, the technology predominantly used to fulfill this need has been developed by the LENA Foundation, which has made advances in the identification and categorization of vocalizations by children and adults~\cite{xu2008signal,gilkerson2017mapping}. Despite these successes, LENA faces significant obstacles to its future viability. Although the LENA software is proprietary and not openly accessible, available descriptions indicate that it still relies on technology that, by 2025, appears to be outdated. Specifically, the software uses a Gaussian Mixture Model with 36 Mel-frequency cepstral coefficients as input, structured within a minimum duration framework and developed under a maximum likelihood estimation approach to categorize audio segments. These segments are classified into broad speaker and nonspeaker categories, which are often simplified into 4 key categories: Key Child, Other Child, Adult Male, and Adult Female, with everything else being treated as background noise or silence. Despite LENA's technological seniority, it remains the benchmark in the field, with anecdotal evidence suggesting that over 95\% of scholarly articles rely on LENA technology.

It was only in 2020 that an open-source alternative emerged, matching or even exceeding LENA's accuracy in the fundamental task of segmenting audio into different speaker types. Capitalizing on the rise of deep learning, Lavechin et al.~\cite{Lavechin2020AnOV} introduced the Voice Type Classifier (VTC), a model that aimed to democratize access to cutting-edge speech processing technology by releasing as open-source the best-performing model, selected after quite extensive experimentation. The released VTC's architecture combined SincNet, to extract low-level features with a stack of LSTMs (Long Short-Term Memory) to aggregate them into context-dependent representations. The best-performing model was trained to return a score for each of Key Child, Other Child, Adult Male, Adult Female, and Speech (a class that was on when any of the others were on, as well as for live speech that human annotators had not been able to assign to a specific source). Unlike LENA, the VTC is freely accessible, updatable, and unconstrained by proprietary recording hardware, making it advantageous for many different use cases (see \cite{lavechin2025performance} for a recent performance comparison).

Despite our excitement about this open-source technology, users and machine learning enthusiasts will be unimpressed by the consideration that VTC outperformed LENA only by a slight margin (F-score of 69\% vs. 55\% for key child; 33\% vs. 29\% for other children; 63\% vs. 43\% for female adult; and 43\% vs. 37\% for male adult, see~\cite{Lavechin2020AnOV} for more details). These results teach us a humbling lesson: detecting \textit{who speaks when} remains vexingly difficult, even for modern deep learning approaches. 

In this paper, we summarize three years of work attempting to improve VTC performance. Since only one previous study \cite{li2024enhancing} investigated foundation models with English long-form data, we contribute novel data using a more diverse dataset and benchmark against the current open-source state-of-the-art \cite{Lavechin2020AnOV}. First, we demonstrate that pre-trained representations learned by Whisper~\cite{radford2023robust} are most effective for smaller training sets. Second, we demonstrate that these pre-trained representations benefit the rare male speech class the most. Third, we provide a detailed analysis of the types of errors made by our model. Finally, we show that there is still room for improvement by benchmarking against human-human agreement. We conclude by reflecting on the numerous directions we explored to improve performance, too many to detail in a 4-page paper.

\section{Methods}

\subsection{End-to-end voice type classification}

As in \cite{Lavechin2020AnOV}, we framed the voice type classification problem as a multi-label classification problem, where the input is the audio stream divided into $N$ frames $S = \{s_1, s_2, \ldots, s_N\}$ and the expected output is the corresponding sequence of labels $\boldsymbol{y} = \{\boldsymbol{y}_1,\boldsymbol{y}_2,\ldots,\boldsymbol{y}_N\}$ where each $\boldsymbol{y}_i$ is of dimension $K$ (the number of labels) with $y_{i,j} = 1$ if the $j^{th}$ class is activated, $y_{i,j} = 0$ otherwise. In training, multiple sequences from the training set were sampled for the total number of frames across all sequences to sum to M in each mini-batch. 

Our baseline model is PyanNet-VTC, PyanNet~\cite{Lavechin2020AnOV,bredin2017pyannote} retrained from scratch. The PyanNet architecture uses a SincNet~\cite{ravanelli2018speaker} feature encoder to learn meaningful filter banks, followed by stacked bidirectional long short-term memory (LSTM) layers and feed-forward (FF) layers.
In contrast, our proposed Whisper-VTC model replaces the SincNet encoder with frozen Whisper representations while introducing a different classification architecture.
For Whisper-VTC, an 80-channel log-magnitude Mel spectrogram is computed using 25-millisecond windows with a 10-millisecond stride. These spectral features are processed through two convolutional layers before entering the frozen Whisper encoder. The encoder outputs are combined using a learnable weighted sum~\cite{xu24c_interspeech}. The key architectural difference is that these representations are then processed by shared bi-LSTM layers (size 256), whose output features are fed to $K=4$ independent binary classification heads, one for each voice type. Each head consists of simple FF layers that output a single scalar for binary classification, unlike PyanNet-VTC's single multi-label output layer.


Due to Whisper's fixed 30-second input requirement and the shorter duration of our audio samples, we truncate the encoder's output representations accordingly. The network is trained to minimize the sum of $K$ independent binary cross-entropy losses:
$$\mathcal{L} = -\dfrac{1}{M} \sum_{i=1}^M \sum_{j=1}^{K}  y_{i,j}\log(\hat{y}_{i,j}) + (1-y_{i,j})\log(1-\hat{y}_{i,j})$$

At test time, we use detection thresholds for each task, by default at $0.5$, as performed during training. These can be tuned to improve performance or balance the precision-recall trade-off post-training.

For our use case, we consider 4 labels: 1) \class{KCHI} for key child vocalizations, 2) \class{OCH} for vocalizations produced by any other children, 3) \class{FEM} for female adult speech, and 4) \class{MAL} for male adult speech. Note that, due to the independent prediction heads, multiple labels can be activated simultaneously in the case of overlapping speech.

The architecture configuration for PyanNet-VTC was the exact same as used in \cite{Lavechin2020AnOV}. 
Whisper-VTC uses bidirectional LSTM layers (size 256) with either 2 or 4 layers showing comparable performance. Table \ref{fig:res1} presents results with 4 LSTM layers.
The last FF stack uses 1 layer of size 256. The learning rate is set up by an AdamW optimizer with a plateau-based learning rate scheduler (for more details, see \mbox{\url{https://github.com/LAAC-LSCP/VTC-IS-25}}).

\subsection{Datasets}

We use the same dataset as in~\cite{Lavechin2020AnOV} that we augmented with three CHILDES \cite{macwhinney1998childes} datasets, which were not collected with a long-form recording device or a wearable but were close enough in domain as to be found to improve performance during piloting (see Table \ref{tab:bbt}). To increase the uncommon category OCH, we additionally segmented sections of audio from the same long-form datasets in BabyTrain-2020 and two new datasets, always following the DARCLE annotation scheme \cite{soderstrom2021developing}. 

\begin{table*}[h]
    \centering
    \caption{Summary statistics of the BabyTrain-2025 dataset. For details about BabyTrain-2020 and the hold-out dataset, see~\cite{lavechin2022reverse}. Most datasets are available via CHILDES~\cite{macwhinney1998childes} or HomeBank~\cite{vandam2016homebank}, which are components of the TalkBank data-sharing platform~\cite{macwhinney2007talkbank}. BabyTrain-2021 highlighted here is a corrected version.
    UK: United Kingdom; NA: North America; PNG: Papua New Guinea}
    \label{tab:bbt}
    \normalsize
    \begin{tabular}{llllrrrr}
        \toprule
         & & & \multicolumn{5}{c}{Cumulated utterance duration} \\
        \cmidrule{5-8}
        Corpus & Access  & Language & Tot. Dur. & \class{KCHI} & \class{OCH} & \class{MAL} & \class{FEM}\\
        \midrule
        BabyTrain-2020~\cite{Lavechin2020AnOV} & Mixture & Mixture & 159h 25m & 39h 58m & 3h 57m & 1h 41m & 64h 45m\\
        BabyTrain-2021 & Mixture & Mixture & 50h 7m & 12h 7m & 4h 8m & 3h 29m & 12h 13m \\
        Forrester~\cite{forrester2002appropriating} & CHILDES & English (UK) & 11h 47m & 4h 28m & 0h 31m & 4h 19m & 2h 29m \\
        Thomas~\cite{lieven2009two} & CHILDES & English (UK) & 403h 18m & 92h 35m & 0h 4m & 0h 57m & 164h 53m \\
        Soderstrom~\cite{soderstrom2008acoustical} & CHILDES & English (NA) & 25h 59m & 2h 30m & 1h 00m & 0h 19m & 12h 30m \\
        Png2019~\cite{cristia2023lena} & Private & Yélî Dnye (PNG) & 0h 23m & 0h 3m & 0h 2m & 0h 1m & 0h 3m\\
        Solomon & Private & Mixture & 5h 23m & 0h 36m & 1h 3m & 0h 32m & 1h 8m\\
        Cougar~\cite{cougar2018} & HomeBank & English (NA) & 13h 0m & 5h 47m & 0h 54m & 1h 25m & 3h 57m\\
        \midrule
        BabyTrain-2025 (total) & Mixture & Mixture & 669h 25m & 158h 8m & 11h 41m &	12h 45m &	262h 0m \\
         \midrule
         Hold-out~\cite{Lavechin2020AnOV} & Mixture & Mixture & 20h 0m & 1h 39m & 0h 45m &	0h 43m &	2h 48m \\
        \bottomrule
    \end{tabular}
    \vspace*{-.5\baselineskip}
\end{table*}

\subsection{Evaluation metrics}

We evaluate PyanNet-VTC and Whisper-VTC using the F-score between precision and recall, the identification error rate and percentage correct, as implemented in \texttt{pyannote.metrics}~\cite{bredin2017pyannote}. 

\section{Results}

\subsection{Experimenting with different Whisper sizes}
\vspace{-15pt}
\begin{figure}[h]
    \centering
    \includegraphics[width=.8\columnwidth]{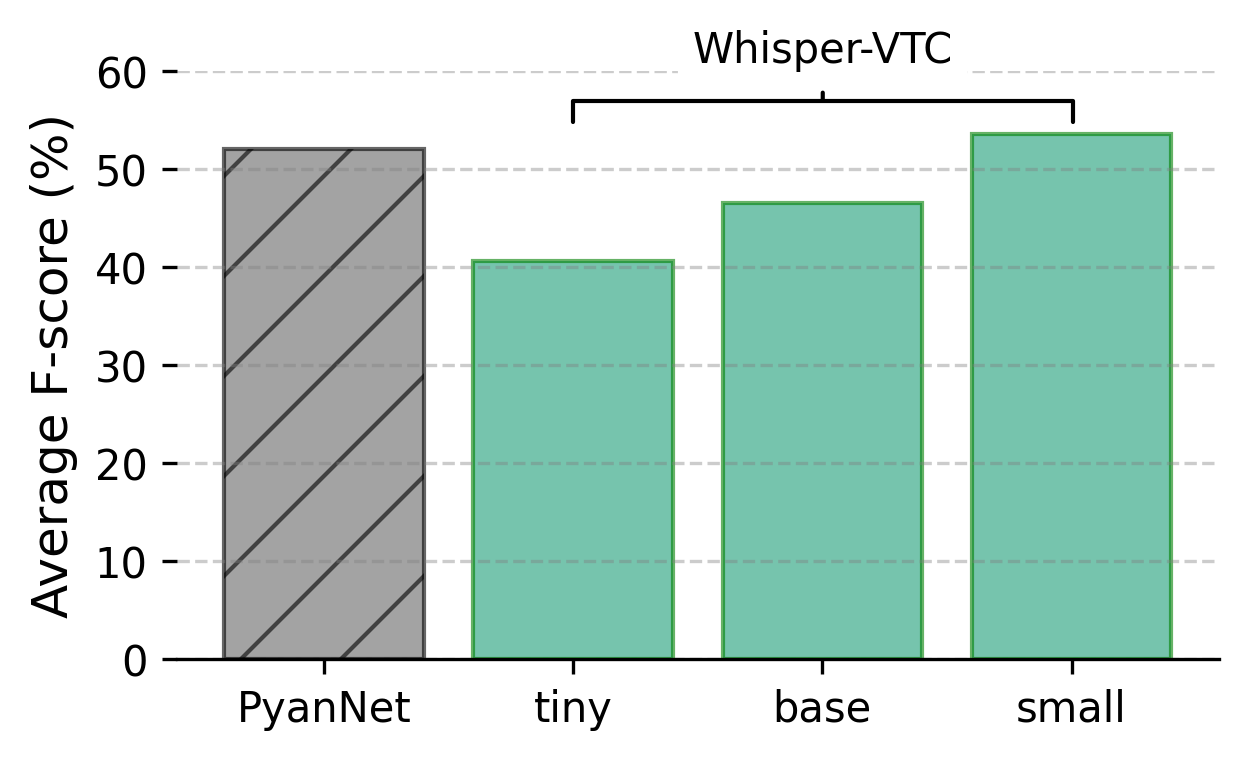}
    \caption{F-score (\%) averaged across speaker categories for PyanNet (results from \cite{Lavechin2020AnOV}) vs. Whisper-VTC (using frozen features from Whisper tiny, base, or small). Performance is computed on the hold-out set.}
    \label{fig:res1}
\end{figure}

Figure \ref{fig:res1} shows a clear trend: larger Whisper-VTC models achieve better performance, with the small model reaching approximately 50\% F-score compared to the tiny model's 40\%. A comparison against PyanNet suggests that using pretrained Whisper representations yields only marginal performance improvement. Additionally, experiments with larger Whisper sizes (not shown here) yield similar performance despite massively longer training times. 

\subsection{Pretrained Whisper representations are most useful for smaller training sets}

\begin{figure}[h]
    \centering
    \includegraphics[width=.82\columnwidth]{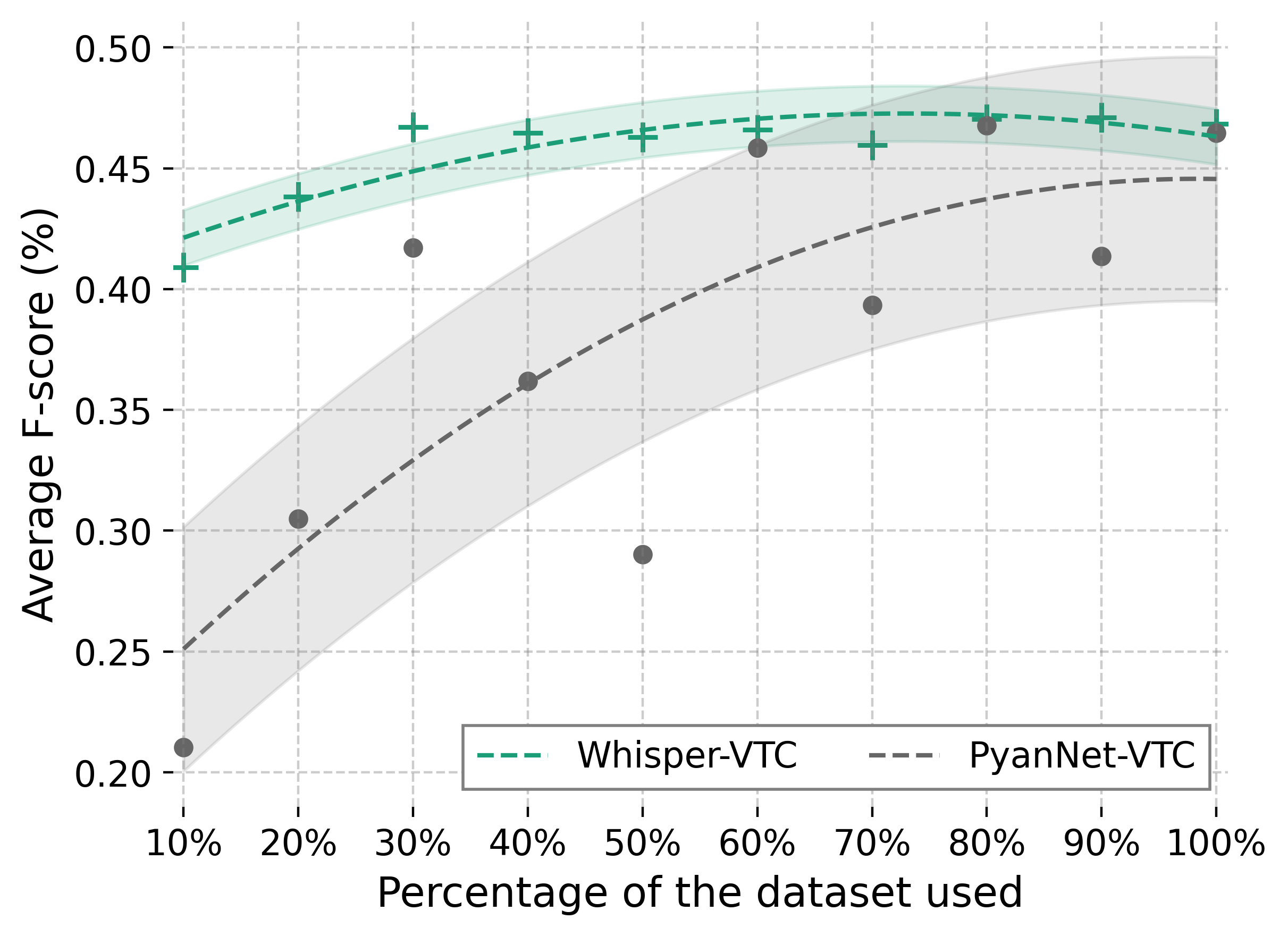}
    \setlength{\belowcaptionskip}{-20pt}
    \caption{F-score (\%) averaged across speaker categories as a function of training set size for  PyanNet-VTC and Whisper-VTC (using frozen features from Whisper base). Performance is computed on the hold-out set.}
    \label{fig:ft4small}
\end{figure}

In this section, we investigate how the performance of Whisper-VTC and PyanNet-VTC varies as a function of training set size.  As evident in Figure \ref{fig:ft4small} (showing only Whisper-VTC base to facilitate inspection), Whisper-VTC yields the most significant advantages compared to PyanNet-VTC when training data is limited. While Whisper-VTC shows markedly superior performance in low-resource scenarios (42\% F-score vs. 35\% with only 10\% of training data), this advantage diminishes as more training data becomes available. Beyond 70\% data utilization (around 420h of audio), both approaches converge to similar average F-scores of approximately 48\%. Interestingly, Whisper-VTC's performance is much more stable across runs than PyanNet-VTC.

\subsection{Segmentation errors: Similarities and differences}

\begin{figure*}
    \centering
    \includegraphics[width=\textwidth]{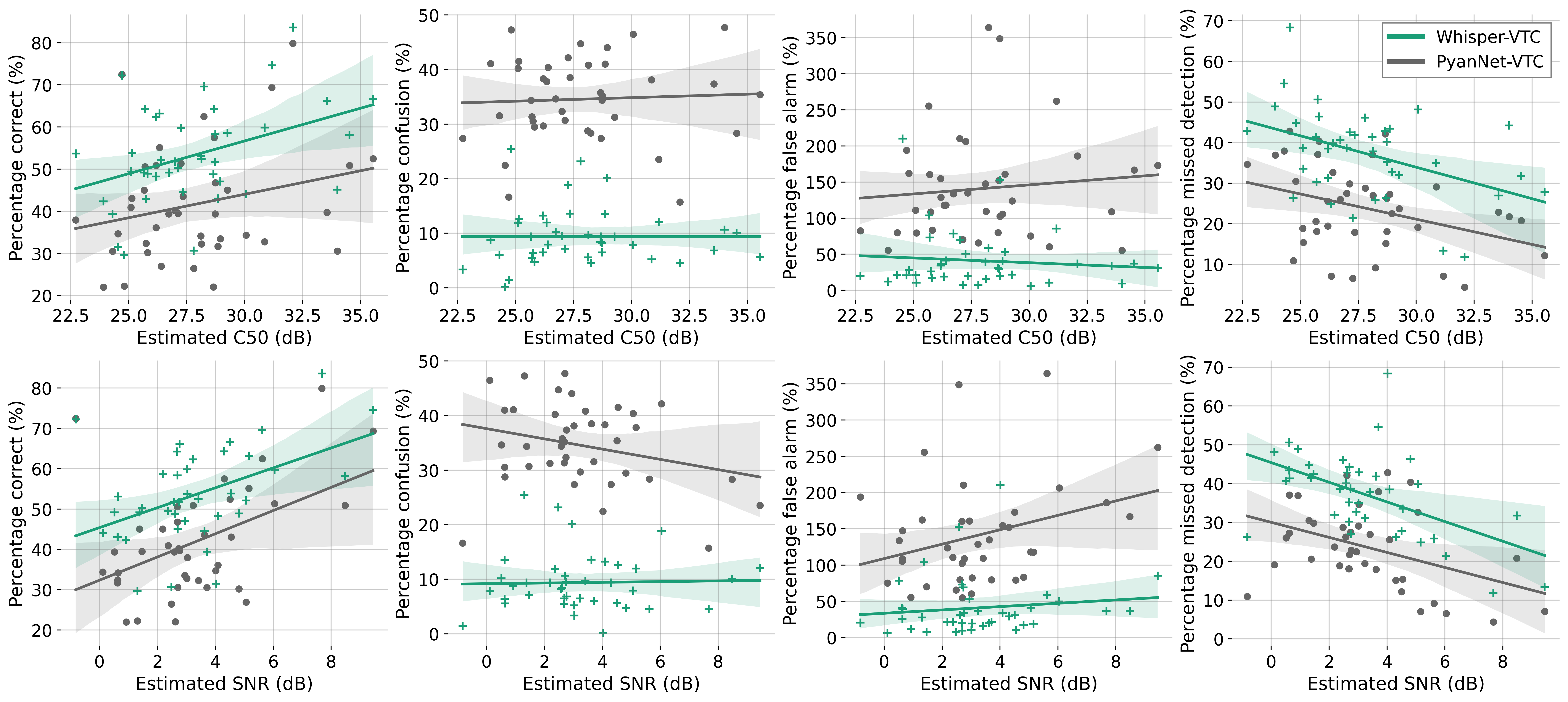}
    \vspace{-10pt}
    \setlength{\belowcaptionskip}{-10pt}
    \caption{Miss (\%), false alarm (\%), confusion (\%), and correct (\%) obtained by PyanNet-VTC and Whisper-VTC (base) as a function of speech-to-noise ratio (SNR) and $C_{50}$ (estimated by Brouhaha~\cite{lavechin2023brouhaha}). Each point represents the average performance over all audio from a given child. Solid lines show linear regression fits with shaded 95\% confidence intervals.}
    \label{fig:seg_errors}
\end{figure*}


Figure \ref{fig:seg_errors} (showing only PyanNet-VTC and Whisper-VTC base to facilitate inspection) suggests that, by and large, Whisper-VTC and PyanNet-VTC behave similarly, presenting higher percent correct and lower miss rates when SNR is higher. Thus, it does not seem like Whisper's extensive pretraining has rendered it more robust to noise than what can be achieved through training with in-domain data. These figures also bring forth a much greater tendency of Whisper-VTC than PyanNet-VTC to miss speech, and for the former to false alarm more than the latter. This difference in behavior is far from obvious, suggesting that perhaps one of the challenges facing systems using Whisper is the ubiquitous presence of far-field speech in long-form recordings. On the other hand, the higher rate of false alarms for PyanNet-VTC is part of the design choice of favoring recall over precision, which could be revisited in the future. 

\subsection{Towards reaching human-level performance}
\label{humanvsmachine}


\begin{table}[htb]
    \centering
    \caption{F-scores (\%) obtained on the hold-out set by PyanNet-VTC~\cite{Lavechin2020AnOV}, Whisper-VTC (using frozen features from Whisper tiny, base or small), and a second human annotator (human 2). Best performances are shown in bold, second best are underlined.}
    \label{tab:comparison}
    \resizebox{\columnwidth}{!}{
    \begin{tabular}{lcccccc}
        \toprule
        System & \class{KCHI} & \class{OCH} & \class{MAL} & \class{FEM} & Ave. \\
        \hline
        Human 2 & \textbf{79.7} & \textbf{60.4} & \textbf{67.6} & \textbf{71.5} & \textbf{69.8} \rule{0pt}{2.6ex} \\
        PyanNet-VTC~\cite{Lavechin2020AnOV} & 68.2 & \underline{30.5} & 41.2 & 63.7 & 50.9 \\
        Whisper-VTC (tiny) & 62,6 & 1.34 & 39.0 & 59.5 & 40.6 \\
        Whisper-VTC (base) & 63.7 & 6.7 & 49.9 & 66.0 & 46.6 \\
        Whisper-VTC (small) & \underline{68.4} & 20.6 & \underline{56.7} & \underline{68.9} & \underline{53.6} \\
        \bottomrule
    \end{tabular}
    }
    \vspace*{-1.2\baselineskip}
\end{table}

Numerous unsuccessful attempts (see Section \ref{sec:disc}) at improving the performance led us to question what performance level was realistically achievable on the voice type classification task.
Table \ref{tab:comparison} shows how our automated systems compare to human performance by presenting F-scores from PyanNet-VTC, Whisper-VTC, and a second human annotator on our hold-out set \cite{soderstrom2021developing}. Note that these recordings are so challenging that even two well-trained human annotators can't agree perfectly, with F-scores as low as 60\% (OCH) and maximally ~80\% (KCHI).  While PyanNet-VTC and Whisper-VTC achieve similar results in the overall averages, both systems underperform compared to human annotation. Focusing on Whisper-VTC, in the best case scenario, the performance gap is just 2\% for \class{FEM}, but still massive for \class{OCH}. This being the most challenging class is reasonable since (a) children are likely under-represented in Whisper's original pretraining data and (b) other children can be far-field in long-form recordings.

\section{Discussion}
\label{sec:disc}
As alluded to in the Introduction, we summarize here some of the extensive experiments we did with null improvements. In one, we found no improvement when replacing the SincNet architecture (PyanNet-VTC) processing raw waveforms with a stack of convolutional layers processing spectrograms. Later, our colleagues raised valid concerns about class imbalance, as Other Children (\class{OCH}) and Male Adult (\class{MAL}) speakers represent only 4.7\% and 1.2\% of the total speech/vocalization time, respectively  -- a well-studied issue~\cite{johnson2019survey,chen2018gradnorm}. To address this, we explored multiple approaches: 1) oversampling the least frequent classes, 2) undersampling the most frequent classes, 3) augmenting the training set with MAL speech from CHiME5~\cite{Barker2018TheF}, and 4) implementing the widely used GradNorm technique to balance gradient norms associated to each class~\cite{chen2018gradnorm}. We also explored powerset transformation of target classes, an approach that reframes multi-label classification into powerset multi-class classification \cite{plaquet23_interspeech}. None of these techniques improved performance~\cite{lucasM2}.
Given our incredibly noisy data, we reasoned that the model could benefit from robust representations learned from massive datasets -- similar to other reports \cite{li2024enhancing}. Our exploration began with wav2vec 2.0 pretrained on LibriLight and our own child-centered long-form recordings before moving to the Whisper experiments reported on in this paper. 

Our novel contribution was not only a new voice-type classifier with careful benchmarking against the most commonly used open-source alternative (PyanNet-VTC), but also the comparison against inter-human agreement to establish the most meaningful performance ceiling. As shown in Section \ref{humanvsmachine}, although Whisper-VTC (small)  outperforms PyanNet-VTC in three classes, and results are particularly encouraging for some classes (notably \class{MAL}), we are still far from human-level performance (particularly for \class{OCH}). The small differences in average F-scores, together with the apparent plateau in Whisper-VTC's performance (Fig. \ref{fig:ft4small}), leave us with little hope for significant improvement in the near future.

Despite sacrificing countless interns' and engineers' time to the altar of performance optimization, all our sophisticated attempts have fallen short of declaring the task of voice type classification ``solved''. These attempts echo Sutton's "bitter lesson"\footnote{The Bitter Lesson of Rich Sutton: \url{http://incompleteideas.net/IncIdeas/BitterLesson.html}}: incorporating domain knowledge through architectural choices and data balancing techniques do not hold a candle to scaling up the amount of data. While it may be tempting to address our performance gap through increasingly sophisticated models or clever data manipulation, our experience suggests that the path forward likely lies in expanding our annotation campaign to build a larger, more diverse training set. This aligns with the historical pattern in AI, where leveraging computation and data quantity has often proven more effective than hand-engineered solutions based on human intuition about the task.

We believe obtaining high-quality, manually annotated data will be crucial. One approach is increased data sharing across research laboratories. The darcle.org mailing list, comprising over 200 researchers working with daylong recordings, exemplifies existing collaborative infrastructure with established data-sharing practices. When we distributed a request for data contributions several months ago, two researchers offered 60 hours of human-segmented audio each, while five others expressed interest but had incompatible sampling or annotation formats. Despite this positive community response, the available contributions would increase our annotated audio corpus by only 14\%, highlighting both the potential and limitations of current data-sharing approaches.


An alternative we hope to explore in future work involves collecting additional data. Messinger et al. \cite{mitsven2022objectively,perry2022reciprocal} have equipped each child and teacher in a classroom with both an audio recording device and RFID-based location tracking (Ubisense Dimension4). This system captures distance and relative angle (i.e., whether individuals are facing each other), offering a promising avenue for automatically generating high-quality labels for key child, female adult, and other child categories. Expanding this approach to home settings could further enable fully automated voice type labeling for all four categories, potentially obviating the need for human annotation.

\section{Conclusions}
We report on over three years of work improving Voice Type Classification in long-form, child-centered recordings—a challenging task due to the realistic recording conditions. Despite these challenges, this technology holds transformative potential for educational interventions and insights into children's language learning mechanisms. Given that errors in voice classification cascade through subsequent analyses, we echo Soderstrom et al. \cite{soderstrom2021developing}'s call for continued collaboration and better-annotated datasets to enhance performance.

\section{Acknowledgements}
This work was performed using HPC resources from GENCI-IDRIS (Grant 2024-AD011015450).

\bibliographystyle{IEEEtran}
\bibliography{mybib}

\end{document}